\documentclass[12pt]{article}
\usepackage[utf8]{inputenc}
\usepackage{graphicx}
\usepackage{titling}
\usepackage[a4paper, left=2.0cm,top=2.0cm,right=2.0cm,bottom=2.0cm]{geometry}
\usepackage{mathptmx}
\usepackage{enumitem}

\usepackage{amsfonts} 
\usepackage{amsmath} 
\usepackage{hyperref}
\usepackage{caption}

\newcommand{\platfont}[1]{\textsc{#1}}

\title{\Large Identification of Cognitive Decline from Spoken Language through Feature Selection and the Bag of Acoustic Words Model}

\author{\normalsize Marko Niemelä$^{1}$, Mikaela von Bonsdorff$^2$, Sami Äyrämö$^1$, and Tommi Kärkkäinen$^1$ 
\vspace{.3cm}\\
1- Faculty of Information Technology, University of Jyväskylä \\
PO Box 35, FI-40014 Jyväskylä, Finland
\vspace{.1cm}\\
2- Faculty of Sport and Health Sciences, University of Jyväskylä \\
PO Box 35, FI-40014 Jyväskylä, Finland
}

\date{}

\begin{document}

\begin{titlingpage}

\maketitle

\begin{abstract}
\normalsize

Memory disorders are a central factor in the decline of functioning and daily activities in elderly individuals. The confirmation of the illness, initiation of medication to slow its progression, and the commencement of occupational therapy aimed at maintaining and rehabilitating cognitive abilities require a medical diagnosis. The earlier a patient receives support and treatment, the better the chances of preserving their ability to work and function. Therefore, the early identification of symptoms of memory disorders, especially the decline in cognitive abilities, plays a significant role in ensuring the well-being of populations. Features related to speech production are known to connect with the speaker's cognitive ability and changes. The lack of standardized speech tests in clinical settings has led to a growing emphasis on developing automatic machine learning techniques for analyzing naturally spoken language. Non-lexical but acoustic properties of spoken language have proven useful when fast, cost-effective, and scalable solutions are needed for the rapid diagnosis of a disease. The work presents an approach related to feature selection, allowing for the automatic selection of the essential features required for diagnosis from the Geneva minimalistic acoustic parameter set and relative speech pauses, intended for automatic paralinguistic and clinical speech analysis. These features are refined into word histogram features, in which machine learning classifiers are trained to classify control subjects and dementia patients from the Dementia Bank's Pitt audio database. The results show that achieving a 75\% average classification accuracy with only twenty-five features with the separate ADReSS 2020 competition test data and the Leave-One-Subject-Out cross-validation of the entire competition data is possible. The results rank at the top compared to international research, where the same dataset and only acoustic features have been used to diagnose patients.

\end{abstract}

\end{titlingpage}

\section{Introduction}

Memory disorder is an illness that impairs memory, information processing, reasoning, and other cognitive functions. Cognitive decline is associated with difficulties in finding and understanding words and interruptions in thoughts \cite{bondi2009neuropsychology}. Dementia is an advanced form of memory disorder where the deterioration of memory and decline in cognitive abilities impair the patient's ability to perform routine daily activities and hinder social and professional functioning. Consequently, advanced dementia has undesirable effects on the patient's immediate circle. The most common cause of progressive memory disorders and dementia is Alzheimer's disease.

According to the World Health Organization (WHO), every third second, someone in the world is diagnosed with a memory disorder \cite{world2017global}. Memory disorders affect over 50 million people worldwide, and the number is expected to triple by the year 2050. The aging of the population mainly influences the prevalence of memory disorders. Consequently, there is a notable increase in countries such as Japan and many European nations where life expectancy has risen, and birth rates have declined. However, memory disorders impact nearly every country worldwide. In the year 2018, the global costs of memory disorders reached nearly a billion dollars \cite{world2017global}.

There is no curative drug treatment for memory disorders, but early diagnosis, treatment, and rehabilitation can improve a patient's functional capacity. Studies have shown that individuals with healthier lifestyles among memory disorder patients experience fewer risk factors for memory disorders and can lead a good and fully functional life \cite{kalaria2008alzheimer}. Diagnosing memory disorders, especially early, has been challenging because diagnostics and treatment evaluation require specialized expertise and monitoring. In recent years, there has been increased investment and focus on developing diagnostics and rehabilitation. The FINGER study was the first in the world to demonstrate that adhering to a comprehensive lifestyle program can improve cognitive functions in the elderly and prevent the decline of memory functions \cite{ngandu20152}.

Cognitive impairment related to memory disorders and potential brain injuries has been studied in clinical settings by analyzing patients' speech production conducted by experts. However, this approach has not proven practical as it is subjective and often challenging to replicate. Consequently, there has been a shift towards developing automatic Natural Language Processing (NLP) machine learning techniques in recent years. These methods have demonstrated utility by being cost-effective, scalable, and providing highly rapid diagnoses \cite{konig2018use} . However, the methods are often too complex, demanding significant training data, computationally intensive, or involve too many adjustable training parameters \cite{roshanzamir2021transformer} \cite{guo2017calibration}.

In the development of machine learning methods, it is crucial to keep the training and test datasets separate to prevent classification results from being biased. In many studies based on machine learning and natural language diagnostics, classification methods are not designed to be independent of patients. Results have been published where multiple recordings from the same patients were used to train and test the machine learning method on data obtained from the same patient. In these cases, classification methods have not learned to identify the disease itself but have instead overfit the speech production characteristics of individual patients, rarely generalizing well to new data. Consequently, classification results on the datasets used in experiments may appear better than they should be \cite{haider2019assessment}.

In machine learning, there has often been a focus on diagnosing Alzheimer's disease or mild cognitive impairment \cite{de2020artificial}. Speech has been analyzed based on lexical, acoustic, or both lexical and acoustic content. Analyzing the lexical content of speech remains challenging because the content needs to be extracted from the recording either manually, which is error-prone and time-consuming, or by using speech-to-text methods that typically require large speech databases and are not language-independent. Additionally, lexical features can easily compromise patient privacy \cite{de2020artificial} .

Hernanz-Dominguez et al. \cite{hernandez2018computer}  conducted a diagnosis based on acoustic features and achieved a classification accuracy of 62.0\% using low-level mel-frequency cepstral coefficients features (MFCC 1-13). The features were extracted from 25 millisecond audio segments, and the following statistical measures were computed: averages, kurtoses, skewness, and variances. In the study \cite{luz2017longitudinal}, memory-impaired patients were diagnosed with an accuracy of 68.0\% based on both the frequencies of speech and speech pauses, including averages, variances, minimum and maximum values, and entropies. In the study \cite{lopez2015automatic}, classification accuracies ranging from 60.0\% to 93.8\% were achieved using speech duration, time- and frequency-domain features, and emotional states. However, the dataset was relatively small (40 individuals). Additionally, the control group included relatively young individuals (20–60 years old) compared to Alzheimer's patients (all over 60 years old).

In the study \cite{LuzSaturninoHai}, classification was performed based on a subset extracted from the Pitt Corpus audio database\footnote{https://dementia.talkbank.org/access/English/Pitt.html}, consisting of control subjects and dementia patients. The database included a total of 156 individuals, half of whom had a diagnosis of dementia. The dataset was divided into training and testing sets based on individuals' diagnoses, ages, and gender distributions. Noise reduction was applied to the audio recordings, and active speech segments were identified using a signal energy-based threshold for speech. Acoustic features were computed for these segments using the functional emobase, computational paralinguistics challenge (compare) \cite{eyben2013recent}, and Geneva minimalistic acoustic parameter set (eGeMAPS) feature sets from the open-source OpenSMILE library\footnote{https://www.audeering.com/research/opensmile/} \cite{eyben2015geneva}. These included low-level descriptors (LLD) and statistical measures for LLD features \cite{eyben2010opensmile}. Additionally, statistical measures were selected for multi-resolution cochleagram features (MRCG), as well as features related to pronunciation and speech pauses used in the study \cite{luz2017longitudinal}. Correlated features were removed from the feature sets. Results for an individual were based on the modes of classification results for all speaker segments using different feature sets. The best classification accuracy based on acoustic features was achieved with the compare feature set. For the test dataset, an individual achieved a prediction accuracy of 62.5\%, and with the LOSO cross-validation used in the experiments, the overall classification accuracy for the entire dataset was 56.5\%.

In the study \cite{syed2020automated}, a classification result of 76.9\% was achieved using LOSO cross-validation on the training data presented in the work \cite{LuzSaturninoHai}. The result is based on acoustic IS10-Paraling features (1582 features sampled from the compare feature set), a bag of acoustic words model (BoAW) \cite{openxbow} provided by the openXBOW tool\footnote{https://github.com/openXBOW/openXBOW}, and support vector machine (linear kernel) and logistic regression classifiers. In the study \cite{haider2019assessment}, a similar data division was utilized as in the study \cite{LuzSaturninoHai}, but with a larger number of selected individuals (a total of 164 individuals). The work introduces a new active data representation method based on self-organizing map (SOM) clustering of feature vectors, associating vectors with their nearest clusters, computing durations of active speech segments, and histogram feature extraction. The proposed model methodologically resembles the BoAW model, where clustering is also used to associate feature vectors with their nearest clusters. The best classification result (77.4\%) is achieved using non-correlated features from the eGeMAPS feature set (a total of 75 features) and a classifier based on linear discriminant analysis (LDA).

The results from studies \cite{haider2019assessment} and \cite{syed2020automated} are promising. However, the papers do not address the challenges of data clustering, such as the SOM network and K-means clustering used in the BoAW model. There are a vast number of ways to partition even a small number of observations into K groups \cite{celebi2013comparative}. Typically, clustering methods converge to a local minimum of the error function, which is why methods should undergo multiple initializations (often one hundred iterations are used \cite{hamalainen2017comparison}\cite{niemela2022improving}), and then the initialization with the smallest clustering error is selected. Multiple initializations do not guarantee an optimal result, but it increases the probability of obtaining a cluster model that effectively describes the internal structure of the data. Another common problem is related to the computational demands of iteratively reassigning clusters in clustering algorithms. In the well-known Lloyd algorithm, the complexity class is $O(nkdi)$, where $n$ is the number of $d$-dimensional input vectors, $k$ is the number of clusters, and $i$ is the number of required iterations \cite{hartigan1979algorithm}. Complexity can be influenced by the choice of the dataset and/or the features.

In this study, feature selection and patient diagnosis based on the BoAW model and classification methods were performed on the ADReSS 2020 competition dataset presented in the study \cite{LuzSaturninoHai}, which includes both control subjects and dementia patients. The study did not use the competition's pre-defined audio segments but utilized complete audio recordings from the competition participants. The recordings underwent data preprocessing, active speech identification, and segmentation. The competition involves two classification tasks. In the first task, a separate test set from the competition, comprising slightly over a third of the individuals in the competition dataset, was classified. In the second task, the generalization of classifiers to the entire competition dataset was measured using LOSO cross-validation. Feature selection was based on the importance scores of a random forest classifier. Feature vectors formed from the most important features were quantized based on clustering and the association of vectors with their nearest cluster centroids. Subsequently, normalized feature histogram representations were created from the quantized vectors used in training and testing six classification models.

\section{Data and Methods}

\subsection{Pitt Corpus audio database}

The Pitt Corpus audio database is one of the databases provided by the DementiaBank\footnote{https://dementia.talkbank.org/access/}. The data for the database was collected between 1983 and 1988 as part of Alzheimer's research at the University of Pittsburgh \cite{becker1994natural}. The study included 282 individuals, out of which 101 were healthy control subjects, and 181 were Alzheimer's patients. Participants had to be over 44 years old, have at least seven years of education, no central nervous system abnormalities, and score at least ten out of thirty points on the Mini-Mental State Examination (MMSE) as a preliminary result \cite{folstein1975mini}.

The selected participants in the study performed oral tasks, and their performance in everyday tasks was also assessed. Among the participants' oral tasks was a kitchen scene description task designed to measure speech disorders (a mother washing dishes, and children standing on stools stealing pastries). A healthcare professional provided instructions to the patients at the beginning of the image description task, and the patients' responses were recorded with a microphone. For the ADReSS 2020 competition\footnote{https://dementia.talkbank.org/ADReSS-2020/}, a random sample of participants in the image description task from the Pitt Corpus database was selected, ensuring that the age and gender distributions of control subjects and dementia patients matched each other \cite{LuzSaturninoHai}. A total of 78 control subjects and 78 dementia patients were chosen. Approximately 69\% of the individuals were selected for the training dataset, and approximately 31\% were selected for the test dataset (Table \ref{tab:1} and Table \ref{tab:2}).

\begin{table}[ht!]
\centering 
\caption{ADReSS 2020 training dataset (M=male, F=female, AD=Alzheimer's disease, MMSE=mini-mental state examination)}
\bigskip
\label{tab:1}       
\begin{tabular}{lllllll}
\hline\noalign{\smallskip}
\multicolumn{4}{c}{AD} & \multicolumn{3}{c}{non-AD} \\
Age & M  & F & MMSE (\emph{SD}) & M & F & MMSE (\emph{SD})   \\
\noalign{\smallskip}\hline\noalign{\smallskip}
{[}50, 55) & 1 & 0 & 30.0 (n/a) & 1 & 0 & 29.0 (n/a) \\
{[}55, 60) & 5 & 4 & 16.3 (4.9) & 5 & 4 & 29.0 (1.3) \\
{[}60, 65) & 3 & 6 & 18.3 (6.1) & 3 & 6 & 29.3 (1.3) \\
{[}65, 70) & 6 & 10 & 16.9 (5.8) & 6 & 10 & 29.1 (0.9) \\
{[}70, 75) & 6 & 8 & 15.8 (4.5) & 6 & 8 & 29.1 (0.8) \\
{[}75, 80) & 3 & 2 & 17.2 (5.4) & 3 & 2 & 28.8 (0.4) \\
\noalign{\smallskip}\hline
Total & 24 & 30 & 17.0 (5.5) & 24 & 30 & 29.1 (1.0) \\
\noalign{\smallskip}\hline
\end{tabular}
\end{table}

\begin{table}[ht!]
\centering 
\caption{ADReSS 2020 test dataset (M=male, F=female, AD=Alzheimer's disease, MMSE=mini-mental state examination)}
\bigskip
\label{tab:2}       
\begin{tabular}{lllllll}
\hline\noalign{\smallskip}
\multicolumn{4}{c}{AD} & \multicolumn{3}{c}{non-AD} \\
Age & M  & F & MMSE (\emph{SD}) & M & F & MMSE (\emph{SD})   \\
\noalign{\smallskip}\hline\noalign{\smallskip}
{[}50, 55) & 1 & 0 & 23.0 (n.a) & 1 & 0 & 28.0 (n.a) \\
{[}55, 60) & 2 & 2 & 18.7 (1.0) & 2 & 2 & 28.5 (1.2) \\
{[}60, 65) & 1 & 3 & 14.7 (3.7) & 1 & 3 & 28.7 (0.9) \\
{[}65, 70) & 3 & 4 & 23.2 (4.0) & 3 & 4 & 29.4 (0.7) \\
{[}70, 75) & 3 & 3 & 17.3 (6.9) & 3 & 3 & 28.0 (2.4) \\
{[}75, 80) & 1 & 1 & 21.5 (6.3) & 1 & 1 & 30.0 (0.0) \\
\noalign{\smallskip}\hline
Total & 11 & 13 & 19.5 (5.3) & 11 & 13 & 28.8 (1.5) \\
\noalign{\smallskip}\hline
\end{tabular}
\end{table}

\subsection{Data pre-processing}

The audio recordings, recorded in WAVE format with a microphone (\emph{Mean}: 55.3s, \emph{SD}: 29.3s), contained a significant amount of noise, which was removed using an adaptive noise reduction filter. The adaptive noise reduction filter introduced short noisy segments at the beginning of the recordings, which were manually removed from the recordings. Additionally, extraneous noises not related to participants, such as caregiver speech, overlapping speech, background noise, buzzer sounds, typing sounds, and emergency vehicle sounds, were removed from the recordings. Towards the end, the recordings only contained either participants' speech or silent periods as participants thought about the task. The high frequencies in the audio recordings did not contain speech, so the recordings were resampled from a 44 kHz sampling frequency to a 16 kHz sampling frequency\footnote{https://sourceforge.net/projects/sox/}, ensuring that the maximum frequency component in each recording was 8 kHz. The resampled audio recordings were normalized to standard audio volumes according to the EBU R 128 Standard\footnote{https://www.mathworks.com/help/audio/ug/loudness-normalization-in-accordance-with-ebu-r-128-standard.html} to minimize the effects of different recording conditions, such as the impact of microphone placement.

\subsection{Active speech recognition}

Segments of active speech were identified from the recordings using the open-source Auditok library\footnote{https://github.com/amsehili/auditok}. The energy threshold for speech recognition (65 dB) and the maximum duration of speech segments (10 seconds) were selected to be the same as in the publication \cite{LuzSaturninoHai}. Default values of the Auditok library were used as limits for the minimum duration of continuous speech (0.2 seconds) and the maximum duration of speech pauses (0.3 seconds). As a result, 2107 audio segments were obtained for the training dataset (\emph{Mean}: 1.27s, \emph{SD}: 0.93s), and 959 audio segments were obtained for the test dataset (\emph{Mean}: 1.28s, \emph{SD}: 0.94s). Finally, the audio segments were normalized to standard audio volumes according to the EBU R 128 Standard.

\subsection{Extraction of audio segment features}

The eGeMAPS provided by the OpenSMILE library is a minimal set of features developed for automatic paralinguistic or clinical speech analysis \cite{eyben2010opensmile}. These features are effective in identifying psychological changes in speech production. They are statistical measures for low-level audio features, such as speech F0 fundamental frequency, speech intensity, spectral flux rate, MFCC features, vibration frequency of the voice (jitter and shimmer), F1, F2, and F3 formant regions, energy ratio (difference between upper and lower spectrum frequencies, i.e., alpha ratio), Hammarberg index, and spectral slope. These features have proven effective in previous studies \cite{eyben2015geneva}. For the audio segments, the OpenSMILE library (\platfont{version 3.0.1}) computed the eGeMAPSv02 features (a total of 88 features). Additionally, the study measured the relative speech pauses before initiating speech production. The speech pause was normalized to the total recording duration and the cumulative speech pause in the recording (a total of two features). Relative speech pauses were added as features of the audio segments.

\subsection{Feature scaling}

Scaling of the data was used in normalizing the feature value scales. Scaling is essential, especially in distance-based applications, because with larger ranges of individual feature variations, computed distances often rely on dominant features. Unlike z-score normalization based on the normal distribution assumption, min-max normalization does not require assumptions about the distributions of features. Min-max scaling to the range $[0, 1]$ corresponds to the following linear transformation:
\begin{equation}
x' = \frac{1}{\max(x)-\min(x)}x + \frac{\min(x)}{\min(x)-\max(x)}, 
\end{equation}
where $x$ is the original variable and $x'$ is the scaled variable. The issue with feature scaling lies in the presence of outliers, which distort normalization parameters. However, it is possible to set the minimum or maximum values of non-outlying data as the values for the outliers.

In the study, during the data preprocessing stage, the features were scaled using min-max scaling as follows:
\ \\
\begin{enumerate}[nolistsep]
\item The training dataset was scaled to the range $[0, 1]$. The normalization parameters were saved.
\item The test dataset was scaled using the normalization parameters from the training data.
\item The features of the test dataset scaled with normalization parameters were required to fall within the range $l = [0 \pm \lambda, 1 \pm \lambda]$, where $\lambda=0.1$. The normalization parameters were fixed for those features that fell within the range $l$. For other features, proceed to step 4.
\item The values of the outlier feature in the training and test datasets were combined into one vector. The step counter was initialized to the value $\beta=0.05$.
\begin{enumerate} 
\item If a feature deviated more than $\lambda$ from the beginning of the range $l$, values in the vector within the range $[0\%, \beta\%)$ were set to be the minimum values in the range $[\beta\%, 100\%]$.
\item If a feature deviated more than $\lambda$ from the end of the range $l$, values in the vector within the range $(100-\beta\%, 100\%]$ were set to be the maximum values in the range $[0\%, 100-\beta\%]$.  
\item The normalization parameters were updated for the feature, and the original non-normalized feature vectors were normalized. If, after normalization, the range of the feature no longer deviated from the test dataset by more than $\lambda$ or $\beta=0.5$, the normalization parameters for the feature were fixed to the new values. Otherwise, the step counter $\beta$ was increased by $\beta=0.05$, and the process returned to step 4.1 for that feature. 
\end{enumerate}
\end{enumerate}

\subsection{Random forest model}

In the feature selection, a random forest model was used, which is a non-linear classification and regression model \cite{breiman2001random}. Random Forest is a meta-estimator that aims to improve the prediction accuracy of a simple decision tree model and prevent overfitting by dividing the data into multiple decision tree estimators and utilizing the average for regression tasks or the mode for classification tasks. The leaf nodes of the random forest trees determine the values of the response variables given the inputs to the tree. In this study, the Gini index of the random forest classifier was used to describe the average impurity decrease, and this can be computed using the probabilities of two or more classes at each tree node. The more the impurity decreases, the more important the feature is in a classification task. The sum of the importance values of the dataset features, computed with the Gini index, is 100\%.

The index is computed as follows:
\begin{equation}
Gini_n = 1 - \sum_{i=1}^kp_{i}^2,  
\end{equation}
where $p_{i}$ represents the probability that samples in node $n$ belong to class $i$ from $k$ classes. In a binary classification task, the formula reads as: 
\begin{equation}
Gini_n = p_1(1-p_1)+p_2(1-p_2) 
\end{equation}

\subsection{Feature selection with random forest}

The utilization of classification models involves selecting only essential features using preprocessed training data cleansed of outliers. Feature selection was conducted as outlined below:
\ \\
\begin{enumerate}[nolistsep]
\item The training dataset was normalized to the $[0, 1]$ scale.
\item The random forest classification model was trained one hundred times on the training dataset. For each iteration, feature importance values were computed based on the Gini index of the random forest.
\item The random forest was trained one hundred times on the training dataset using permuted response variables. For each iteration, permuted importance values were computed for the features. 
\item The distributions of one hundred importance values for each feature and the importance values obtained through permutations were compared using the statistical Wilcoxon signed-rank test. The null hypothesis assumed that the values between the groups come from the same median distribution at a 5\% statistical significance level ($p=0.05$). 
\item The final features were selected as the k highest-ranked features in terms of average importance values that rejected the null hypothesis of the Wilcoxon test.
\end{enumerate}

\subsection{Bag of acoustic words model}

The word histogram model represents an object as an unordered collection of words. The model has been used in text categorization \cite{joachims1998text}, such as spam filtering. The word histogram model has been expanded to include visual words (bag-of-visual-words), used in image categorization \cite{csurka2004visual}, and acoustic words (bag-of-acoustic-words), for example, in natural language processing of spoken language \cite{weninger2013words}. Typically, the content of speech is not directly used in the model. Instead, compact feature vector representations are derived from speech segments, for instance, by analyzing short-term frequency spectra. Following this, a histogram distribution of feature vectors is constructed, which is then utilized in analyzing the speech.

\begin{figure}[ht!]
\centering	
  \includegraphics[width=1.0\textwidth]{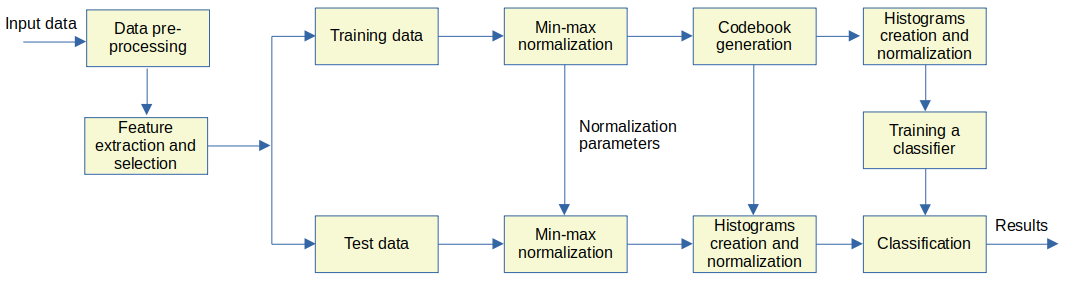}
\caption{Classification process over bag of acoustic word model.}
\label{fig:1}      
\end{figure}

In Figure \ref{fig:1}, the employed word histogram method is illustrated. Initially, for the word histogram model, preprocessing, feature extraction, and feature scaling were applied to the training and test data. Subsequently, a limited number of cluster centroids, or cluster prototypes, were formed from the feature vectors of the training dataset by separately clustering the data of control subjects and dementia patients into a specified number of clusters representing the two different classes of the training data. K-means, one of the well-known clustering methods \cite{jain2010data}, is not robust to outliers because cluster prototypes are represented by the averages of the data sets. The K-spatial median clustering used in the experiments, combined with an appropriate initialization method, is robust because the prototypes are based on the multidimensional medians of the sets \cite{ayramo2006knowledge}\cite{ayramo2007robust}. The clustering initialization was performed a total of one hundred times using the K-means++ method, which randomly selects starting points and favors points far from each other \cite{arthur2007k}. The clustering result selected was the outcome of the initialization that best minimized the clustering error based on Euclidean distance:
\begin{equation}
\label{eq:1}   
J = \sum_{k=1}^K\sum_{i=1}^N\|\mathbf{x}_i-\mathbf{c}_k\|_2
\end{equation}
where $\mathbf{X} = \{\mathbf{x}_i\}_{i=1}^N$ denotes data set and $\{\mathbf{c}_k\}_{k=1}^K$ is the set of cluster prototypes that minimizes locally the error function (See eq. (\ref{eq:1})). 

A collection of prototype vectors (codebook) was formed from clustered feature vector representations. The feature vectors of speech segments were quantized by associating the vectors with the nearest prototype vectors in the codebook (words) based on Euclidean distances. The entire content of the speech was described by a histogram, where each bin represented the occurrence of a codebook word in the speech. After forming the histogram, it was normalized by the total number of words in the histogram. Normalized histograms were used in classification.

\subsection{LOSO cross-validation}
\label{Sec:LOSO}

In Leave-One-Subject-Out (LOSO) cross-validation, all $N$ observations in the dataset are divided into $N$ sets. In each iteration, $N-1$ sets are used for training the model, and the remaining one set is used for testing the model. The validation process is repeated $N$ times, ensuring that each observation is tested exactly once.

\subsection{Comparison of classification methods}

Six different methods were used for classification: five-nearest neighbors classifier (5-NN), LDA, random forest (50 decision trees and 5 leaf nodes), extreme minimal learning machine classifier (reference point rate $RPRat=100.0\%$), linear support vector machine (cost parameter $C=1.0$), and support vector machine with chi-squared kernel (cost parameter $C=0.25$). Among the classifiers, 5-NN, extreme minimal learning machine (EMLM), and linear support vector machine (Linear-SVM) were based on MATLAB implementations (\platfont{version R2022b, 64-bit}), LDA and random forest (RF) were based on the scikit-learn library in Python (\platfont{version 1.2.2}), and support vector machine with chi-squared kernel (Chi2-SVM) was based on the libsvm library (\platfont{version 3.32}).

The support vector machine with a chi-squared kernel has proven to be effective in histogram classification \cite{zhang2007local}. The chi-squared kernel is based on chi-squared distances, which are incorporated into the kernel function using an extended Gaussian kernel:
\begin{equation}
K \left( S_i, S_j \right) = exp \left(\frac{-1}{A}D\left( S_i, S_j\right)\right), 
\end{equation}
where, $D(S_i, S_j)$ represents the chi-squared distance between word histograms $S_i$ and $S_j$, and $A$ denotes a scaling parameter, which is the average of all training histogram chi-squared distances.

The performance of the classification models was evaluated using the ADReSS 2020 competition test dataset (48 individuals) and by conducting LOSO cross-validation on the entire competition dataset (156 individuals). The same classifier parameters were utilized in both experiments.

\section{Results}

\subsection{Feature selection}

\begin{figure}[ht!]
  \includegraphics[width=0.55\textwidth]{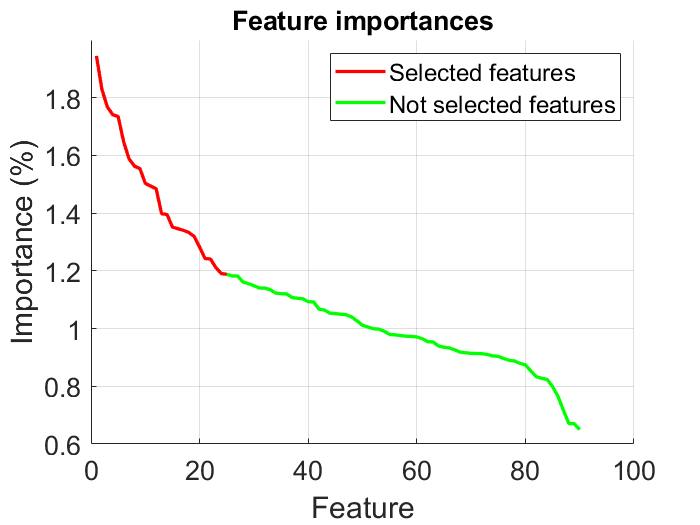}
\caption{Average feature importances in sorted order for eGeMAPS feature set and relative speech pauses.}
\label{fig:2}      
\end{figure}

In the feature selection based on the Gini index and the Wilcoxon signed-rank test for the random forest classifier, a total of 25 features were identified. Among the most important features were functional features related to MFCC 1-4 coefficients, speech F0 fundamental frequency, spectral slopes (slopeV0-500 and slopeV500-1500), loudness, harmonic features (H0, H1, and H2), power spectrum (spectralFlux), and F3 formant region. Additionally, important features included normalized speech pauses (pauseTotalDurationRatio and pauseTotalPausesRatio). The combined sum of the average importance scores for the selected features was 36.7\%. Figure \ref{fig:2} illustrates the average importance scores of selected and non-selected features. All eGeMAPS features and their average importance scores are listed in Appendix A.

\subsection{Classification of test data}

The performance of classifiers was measured with selected twenty-five features and different-sized codebooks. The cluster prototypes of control subjects and dementia patients were both increased in increments of five from five prototypes to fifty prototypes. Table 3 illustrates the results for a hundred iterations of clustering, each with a hundred clustering reinitializations. From the table, it can be observed that the Chi2-SVM performed the classification task best, with an average classification accuracy of 75.2\% (+/- 4.0\%) on the test dataset. The RF model used for feature selection achieved a good classification accuracy of 71.9\% (+/- 4.8\%). The best results were obtained with a codebook of thirty prototypes. The variability in classification accuracies can be explained in part by the randomness associated with classification models and partly by the randomness in the selection of clustering starting points. The variation in 5-NN classification results (2.9\%-5.3\%) is entirely explained by the randomness of starting points since the classification model itself does not involve randomness. Table \ref{tab:3} indicates that the performance of classifiers begins to decline as the codebook size increases beyond thirty prototypes.

\begin{table}[ht!]
\caption{Classification accuracies and standard deviations for ADReSS 2020 test data set over 100 repetitions of replicated clustering}
\label{tab:3}       
\begin{tabular}{llllllll}
\hline\noalign{\smallskip}
 & Classif. & 5-NN & RF & Linear-SVM & Chi2-SVM & LDA & EMLM \\
NClust. & & & & & & & \\
\hline\noalign{\smallskip}
5+5 & & 68.9\% & 63.7\% & 64.3\% & 59.6\% & 60.6\% & 57.5\% \\
 & & ($\pm$ 2.9\%) & ($\pm$ 4.5\%) & ($\pm$ 1.7\%) & ($\pm$ 2.1\%)& ($\pm$ 2.5\%) & ($\pm$ 2.9\%) \\
10+10 & & \textbf{70.2\%} & 71.5\% & 66.9\% & 70.2\% & \textbf{67.7\%} & 66.3\% \\
 & & \textbf{($\pm$ 5.0\%)} & ($\pm$ 4.3\%) & ($\pm$ 3.3\%) & ($\pm$ 3.2\%)& \textbf{($\pm$ 3.9\%)} & ($\pm$ 3.7\%) \\
15+15 & & 66.9\% & \textbf{71.9\%} & \textbf{68.4\%} & \textbf{75.2\%} & 64.5\% & \textbf{67.3\%} \\
 & &  ($\pm$ 5.2\%) & \textbf{($\pm$ 4.8\%)} & \textbf{($\pm$ 4.5\%)} & \textbf{($\pm$ 4.0\%)} & ($\pm$ 5.3\%) & \textbf{($\pm$ 4.6\%)} \\
20+20 & & 64.9\% & 70.7\% & 66.3\% & 73.7\% & 62.0\% & 65.3\% \\
 & &  ($\pm$ 4.5\%) & ($\pm$ 4.7\%) & ($\pm$ 3.9\%) & ($\pm$ 3.8\%)& ($\pm$ 6.0\%) & ($\pm$ 5.3\%) \\
25+25 & & 64.7\% & 68.5\% & 64.4\% & 70.8\% & 61.0\% & 63.4\% \\
 & &  ($\pm$ 4.1\%) & ($\pm$ 4.4\%) & ($\pm$ 3.8\%) & ($\pm$ 4.1\%)& ($\pm$ 5.5\%) & ($\pm$ 5.0\%) \\
30+30 & & 63.7\% & 67.9\% & 64.3\% & 68.8\% & 59.8\% & 63.0\% \\
 & & ($\pm$ 4.6\%) & ($\pm$ 4.5\%) & ($\pm$ 3.8\%) & ($\pm$ 4.2\%)& ($\pm$ 5.6\%) & ($\pm$ 5.0\%) \\
35+35 & & 63.3\% & 68.2\% & 63.7\% & 68.9\% & 59.3\% & 63.6\% \\
 & &  ($\pm$ 4.6\%) & ($\pm$ 5.4\%) & ($\pm$ 4.5\%) & ($\pm$ 4.4\%)& ($\pm$ 5.9\%) & ($\pm$ 4.9\%) \\
40+40 & & 62.8\% & 66.5\% & 63.1\% & 67.9\% & 57.2\% & 61.9\% \\
 & & ($\pm$ 4.9\%) & ($\pm$ 5.3\%) & ($\pm$ 4.4\%) & ($\pm$ 4.1\%)& ($\pm$ 6.1\%) & ($\pm$ 4.8\%) \\
45+45 & & 61.8\% & 65.8\% & 62.5\% & 67.4\% & 55.0\% & 62.7\% \\
 & &  ($\pm$ 4.6\%) & ($\pm$ 5.4\%) & ($\pm$ 4.6\%) & ($\pm$ 4.6\%)& ($\pm$ 5.4\%) & ($\pm$ 4.1\%) \\
50+50 & & 62.2\% & 65.0\% & 61.9\% & 66.8\% & 54.3\% & 62.6\% \\ 
 & & ($\pm$ 5.3\%) & ($\pm$ 5.8\%) & ($\pm$ 4.8\%) & ($\pm$ 4.6\%)& ($\pm$ 7.2\%) & ($\pm$ 4.9\%) \\
\noalign{\smallskip}\hline
\end{tabular}
\end{table}

\subsection{Performance in LOSO cross-validation}

\begin{figure}[ht!]
\centering	
  \includegraphics[width=1.0\textwidth]{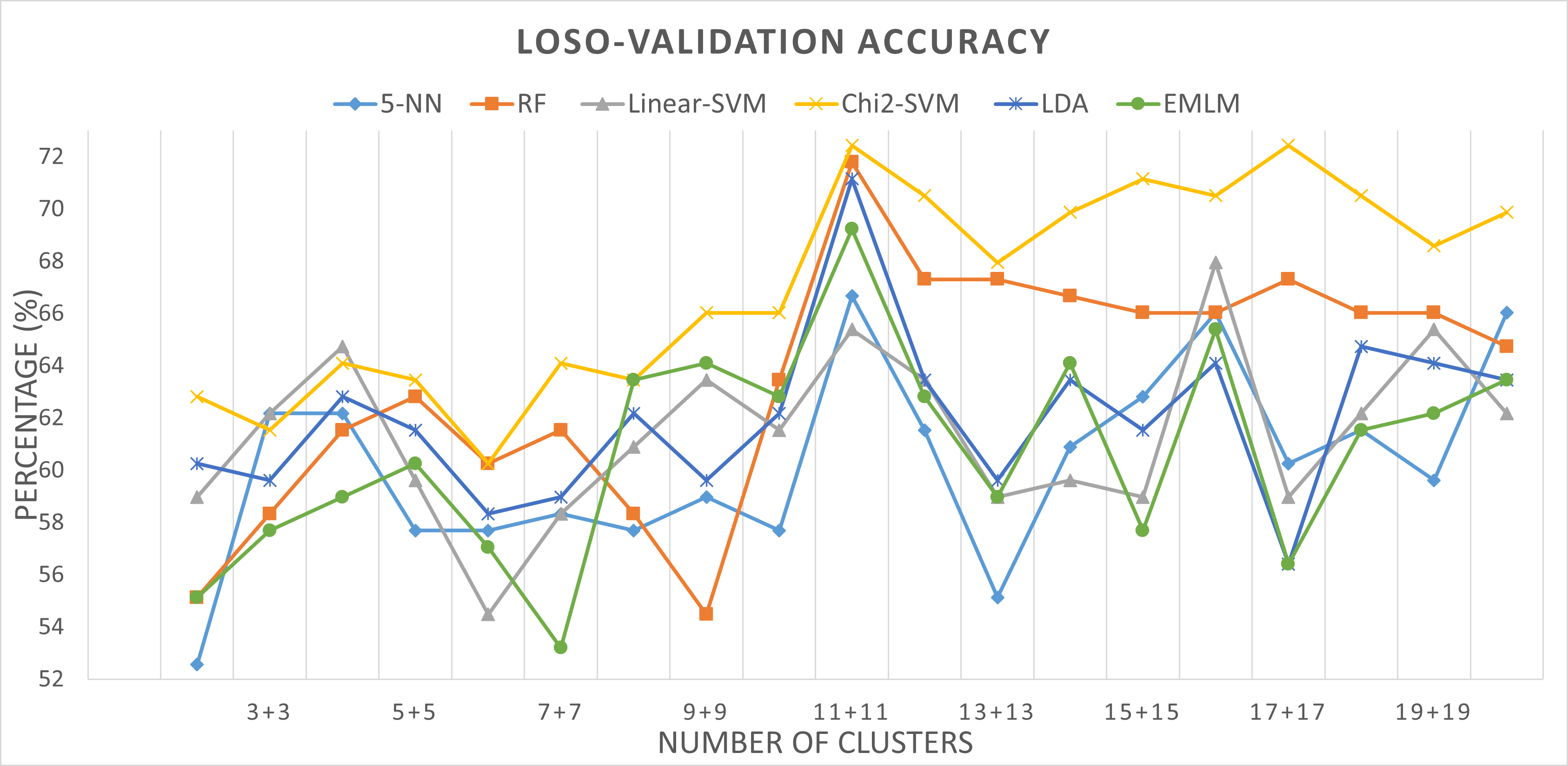}
\caption{LOSO cross-validation accuracy over 25 repetitions of replicated clustering.}
\label{fig:3}      
\end{figure}

The generalization performance of classification models was measured by LOSO cross-validation on the dataset of 156 individuals from the ADReSS 2020 competition. The size of the codebook was increased from four cluster prototypes to forty prototypes. The uncertainty associated with cluster error was addressed by repeating LOSO validations a total of twenty-five times, with each clustering iteration involving a hundred reinitializations. The results for each individual were obtained by considering modes and the smallest clustering errors from the twenty-five classification outcomes. Figure \ref{fig:3} illustrates the classification results for the entire dataset using different classifiers, based on either modes or the smallest clustering errors (the better of the two results was always chosen). From Figure \ref{fig:3}, it is evident that the best results for most classifiers were achieved with a codebook size of twenty-two prototypes. The Chi2-SVM classifier performed the classification task best, with a classification accuracy of 72.4\%. Additionally, RF and LDA produced classification results of over 70\%. Table \ref{tab:4} provides LOSO validation results with a codebook size of twenty-two prototypes and seventy-five clustering iterations. The classification results are based on the smallest clustering error outcome. Clearly, the Chi2-SVM classifier emerged as the top performer in LOSO validation with an accuracy of 75.0\%. Among the classifiers, EMLM and RF achieved classification accuracies of 71.2\% and 70.5\%, respectively.

\begin{table}[ht!]
\centering 
\caption{LOSO cross-validation results based on 22 cluster prototypes and 75 repetitions of replicated clustering}
\bigskip
\label{tab:4}       
\begin{tabular}{llcccccc}
\hline\noalign{\smallskip}
 & Classif. & 5-NN & RF & Linear-SVM & Chi2-SVM & LDA & EMLM \\
NClust. & & & & & & & \\
\hline\noalign{\smallskip}
 11+11 & & 64.1\% & 70.5\% & 66.0\% & \textbf{75.0\%} & 69.9\% & 71.2\% \\
\noalign{\smallskip}\hline
\end{tabular}
\end{table}

\section{Discussion}

In this study, diagnostic analysis related to dementia was conducted on a dataset extracted from the Pit Corpus audio database, consisting of both control subjects and dementia patients. Statistical features for acoustic speech analysis were extracted from audio segments using the OpenSMILE library, and histograms were constructed based on the BoAW approach. Similar analyses, based on the same dataset and a comparable histogram representation, have been conducted in previous studies, with reported LOSO cross-validation accuracies of 76.9\% and 77.4\% \cite{haider2019assessment}\cite{syed2020automated}. However, these studies do not provide details on how they accounted for randomness in the clustering process employed for histogram formation.

The study experiments with the ADReSS 2020 competition test dataset reveal that, through a hundred iterations of clustering, the variability in classification outcomes using a deterministic 5-NN classifier typically falls within the range of 4.0\% to 5.0\%. Therefore, in the experiments, twenty-five repetitions of LOSO cross-validation were employed when searching for the number of clusters, and seventy-five repetitions were used in the final results computation. It is important to note that the LOSO validation method itself involves repetitions of grouping equal to the number of subjects to be validated, leading to a multiplication of the final repetitions by the number of subjects (see Section \ref{Sec:LOSO}).

In the study \cite{syed2020automated}, a large set of acoustic features (a total of 1582 features) has been used, leading to computational complexity for the methods. In this study, the computation of average feature importance scores was performed for relative speech pauses and eGeMAPS features (a total of 90 features) using a random forest classifier based on the Gini index. Feature selection was conducted based on the null hypothesis of the Wilcoxon statistical test. Feature selection relying on statistical testing is a straightforward and relatively efficient method for selecting final features \cite{wilcoxon1992individual}.

In literature, feature selection has been performed for the eGeMAPS feature set in emotion recognition based on speech recordings \cite{haider2021emotion}. The study utilized four different feature selection mechanisms. In the Infinite Latent Feature Selection (ILFS) method, feature importance values are based on all possible subsets of features, representing different paths in the feature graph. The ReliefF method uses weight vectors to represent the connections of features to actual class labels. The Generalized Fisher Score method (Fisher) seeks a set of features that maximizes the lower bound of a credibility function based on the Fisher metric. The Active Feature Selection method (AFS) clusters individual features of the dataset and selects final features based on the distinctiveness of clusters. In the study \cite{haider2021emotion}, three datasets related to emotion recognition were combined into one set, and the feature selection methods proposed 44–79 eGeMAPS features as the final quantities of features.

In this study, among the key features, statistical features of the MFCC and statistical features related to the fundamental frequency (F0) of the voice were particularly prominent. Previous studies have used low-level MFCC features together with signal energy for emotion recognition \cite{schmitt16_interspeech}. Additionally, the fundamental frequency of the voice is known to be associated with Parkinson's disease \cite{sidtis2004fundamental}, and the fundamental frequency of vowel sounds, along with frequency oscillation, has been investigated in the diagnosis of Parkinson's disease \cite{little2008suitability}. In the study \cite{schmitt16_interspeech}, instead of active speech recognition, features were extracted from unsegmented audio recordings in 25 ms durations using a 10 ms audio sampling. For low-level features, statistical features (means and standard deviations) were also computed, and the obtained classification results were compared. The presented previous research results support the broader scalability of the features and methods used in this study to other speech-identifiable common diseases or different emotions, which could serve as indicators for illnesses or depression \cite{dimsdale2008psychological}\cite{desmet2013emotion}.

\section{Conclusion}

The goal of this study was to attempt to identify individuals with cognitive decline based on naturally spoken language. The study did not aim for challenging speech recognition, which could involve transforming speech into text and interpreting the content of the produced text. Instead, spoken language was categorized based on the acoustic characteristics of the language. In the study, audio recordings were segmented into at least 20 ms long audio segments, and features were computed for the segments using the open-source OpenSMILE library. Additionally, relative speech pauses occurring in speech were added to the features of speech segments.The BoAW model was used in speech diagnosis, where the feature vectors of segments were quantized based on iterations of K-spatial-medians clustering and using a codebook of acoustic words. A random forest model and feature selection related to statistical testing were employed to reduce the computational complexity of the BoAW model and classification methods. Based on the results, it is possible to classify control individuals and dementia patients simply, reliably, and efficiently with an average accuracy of 75.0\% using only twenty-five acoustic features, emphasizing MFCC (1–4) and statistical characteristics of the fundamental frequency (F0). Low-level descriptions of features are known to be related to various emotions and illnesses. The low number of key features provides researchers with opportunities to further develop feature extraction in a more reliable direction. For instance, it is possible to compute new and more efficient statistics for key features based on statistical methods and correlation testing. Additionally, low-level features can be directly applied to unsegmented recordings using a fixed-size sampling window \cite{schmitt16_interspeech}. Acoustic features are independent of the content of speech. Therefore, the developed diagnostic methods can be expanded to multiple languages based on openly available audio databases.

\section*{Acknowledgement}

The first author would like to express gratitude to the Finnish Cultural Foundation for the financial support.

%
%
%
%
%

\bibliographystyle{ieeetr}
\bibliography{mybib}

\clearpage

\appendix
\section{Average importances for eGeMAPS features and relative speech pauses}

\captionsetup[table]{labelformat=empty}
\begin{table}[ht!]
\footnotesize
\centering 
\caption{Average importances for eGeMAPS feature set and speech pauses based on Gini index (first 25 features are selected to the study)}
\vspace{10px}
\label{FeatImp}       
\begin{tabular}{lll}
\hline\noalign{\smallskip}
Rank & Feature & Importance (\%)  \\
\noalign{\smallskip}\hline\noalign{\smallskip}
1 & \emph{mfcc3V\_sma3nz\_amean} & \textbf{1.95} \\
2 & \emph{logRelF0-H1-A3\_sma3nz\_amean} & \textbf{1.83} \\
3 & \emph{mfcc3\_sma3\_amean} & \textbf{1.77} \\
4 & \emph{slopeV0-500\_sma3nz\_amean} & \textbf{1.74} \\
5 & \emph{F0semitoneFrom27.5Hz\_sma3nz\_percentile50.0} & \textbf{1.73} \\
6 & \emph{F0semitoneFrom27.5Hz\_sma3nz\_percentile20.0} & \textbf{1.65} \\
7 & \emph{loudness\_sma3\_amean} & \textbf{1.59} \\
8 & \emph{slopeUV0-500\_sma3nz\_amean} & \textbf{1.56} \\
9 & \emph{F3amplitudeLogRelF0\_sma3nz\_stddevNorm} & \textbf{1.55} \\
10 & \emph{loudness\_sma3\_percentile20.0} & \textbf{1.5} \\
11 & \emph{F0semitoneFrom27.5Hz\_sma3nz\_percentile80.0} & \textbf{1.49} \\
12 & \emph{F0semitoneFrom27.5Hz\_sma3nz\_amean} & \textbf{1.48} \\
13 & \emph{mfcc4\_sma3\_amean} & \textbf{1.4} \\
14 & \emph{logRelF0-H1-H2\_sma3nz\_amean} & \textbf{1.4} \\
15 & \emph{mfcc2\_sma3\_amean} & \textbf{1.35} \\
16 & \emph{mfcc4V\_sma3nz\_amean} & \textbf{1.35} \\
17 & \emph{F3bandwidth\_sma3nz\_amean} & \textbf{1.34} \\
18 & \emph{slopeV0-500\_sma3nz\_stddevNorm} & \textbf{1.33} \\
19 & \emph{spectralFluxUV\_sma3nz\_amean} & \textbf{1.32} \\
20 & \emph{pauseDurationRatio} & \textbf{1.28} \\
21 & \emph{mfcc1V\_sma3nz\_amean} & \textbf{1.24} \\
22 & \emph{slopeV500-1500\_sma3nz\_amean} & \textbf{1.24} \\
23 & \emph{loudness\_sma3\_percentile50.0} & \textbf{1.21} \\
24 & \emph{pauseTotalPausesRatio} & \textbf{1.19} \\
25 & \emph{mfcc2V\_sma3nz\_amean} & \textbf{1.19} \\
26 & slopeUV500-1500\_sma3nz\_amean & 1.18 \\
27 & mfcc2V\_sma3nz\_stddevNorm & 1.18 \\
28 & loudness\_sma3\_pctlrange0-2 & 1.16 \\
29 & slopeV500-1500\_sma3nz\_stddevNorm & 1.16 \\
30 & loudness\_sma3\_stddevNorm & 1.15 \\
31 & HNRdBACF\_sma3nz\_amean & 1.14 \\
32 & mfcc1\_sma3\_amean & 1.14 \\
33 & mfcc3\_sma3\_stddevNorm & 1.13 \\
34 & loudness\_sma3\_percentile80.0 & 1.12 \\
35 & hammarbergIndexV\_sma3nz\_amean & 1.12 \\
36 & shimmerLocaldB\_sma3nz\_amean & 1.12 \\
37 & alphaRatioUV\_sma3nz\_amean & 1.11 \\
38 & hammarbergIndexUV\_sma3nz\_amean & 1.11 \\
39 & logRelF0-H1-H2\_sma3nz\_stddevNorm & 1.1 \\
40 & mfcc4\_sma3\_stddevNorm & 1.09 \\
41 & F3bandwidth\_sma3nz\_stddevNorm & 1.09 \\
42 & shimmerLocaldB\_sma3nz\_stddevNorm & 1.07 \\
43 & spectralFlux\_sma3\_amean & 1.06 \\
44 & mfcc4V\_sma3nz\_stddevNorm & 1.05 \\
45 & mfcc1\_sma3\_stddevNorm & 1.05 \\
\noalign{\smallskip}\hline
\end{tabular}
\end{table}

\captionsetup[table]{labelformat=empty}
\begin{table}[ht!]
\centering 
\footnotesize
\caption{Average importances for eGeMAPS feature set and speech pauses based on Gini index}
\vspace{10px}
\label{FeatImp:2}       
\begin{tabular}{lll}
\hline\noalign{\smallskip}
Rank & Feature & Importance (\%)  \\
\noalign{\smallskip}\hline\noalign{\smallskip}
46 & mfcc3V\_sma3nz\_stddevNorm & 1.05 \\
47 & equivalentSoundLevel\_dBp & 1.05 \\
48 & F2bandwidth\_sma3nz\_amean & 1.04 \\
49 & F0semitoneFrom27.5Hz\_sma3nz\_pctlrange0-2 & 1.03 \\
50 & F3frequency\_sma3nz\_amean & 1.01 \\
51 & F2amplitudeLogRelF0\_sma3nz\_stddevNorm & 1.01 \\
52 & mfcc1V\_sma3nz\_stddevNorm & 1.00 \\
53 & mfcc2\_sma3\_stddevNorm & 1.00 \\
54 & spectralFlux\_sma3\_stddevNorm & 0.99 \\
55 & F1frequency\_sma3nz\_amean & 0.98 \\
56 & F1frequency\_sma3nz\_stddevNorm & 0.98 \\
57 & HNRdBACF\_sma3nz\_stddevNorm & 0.98 \\
58 & logRelF0-H1-A3\_sma3nz\_stddevNorm & 0.97 \\
59 & jitterLocal\_sma3nz\_amean & 0.97 \\
60 & F0semitoneFrom27.5Hz\_sma3nz\_stddevNorm & 0.97 \\
61 & F2frequency\_sma3nz\_amean & 0.97 \\
62 & jitterLocal\_sma3nz\_stddevNorm & 0.95 \\
63 & loudness\_sma3\_meanFallingSlope & 0.95 \\
64 & F1bandwidth\_sma3nz\_stddevNorm & 0.94 \\
65 & loudness\_sma3\_stddevRisingSlope & 0.94 \\
66 & MeanVoicedSegmentLengthSec & 0.93 \\
67 & loudness\_sma3\_meanRisingSlope & 0.93 \\
68 & F1amplitudeLogRelF0\_sma3nz\_stddevNorm & 0.92 \\
69 & spectralFluxV\_sma3nz\_amean & 0.92 \\
70 & alphaRatioV\_sma3nz\_amean & 0.91 \\
71 & F3frequency\_sma3nz\_stddevNorm & 0.91 \\
72 & F0semitoneFrom27.5Hz\_sma3nz\_meanRisingSlope & 0.91 \\
73 & F1bandwidth\_sma3nz\_amean & 0.91 \\
74 & F2frequency\_sma3nz\_stddevNorm & 0.91 \\
75 & F2bandwidth\_sma3nz\_stddevNorm & 0.9 \\
76 & hammarbergIndexV\_sma3nz\_stddevNorm & 0.9 \\
77 & F0semitoneFrom27.5Hz\_sma3nz\_meanFallingSlope & 0.89 \\
78 & alphaRatioV\_sma3nz\_stddevNorm & 0.89 \\
79 & spectralFluxV\_sma3nz\_stddevNorm & 0.88 \\
80 & loudnessPeaksPerSec & 0.87 \\
81 & VoicedSegmentsPerSec & 0.85 \\
82 & F3amplitudeLogRelF0\_sma3nz\_amean & 0.83 \\
83 & MeanUnvoicedSegmentLength & 0.83 \\
84 & loudness\_sma3\_stddevFallingSlope & 0.82 \\
85 & F2amplitudeLogRelF0\_sma3nz\_amean & 0.8 \\
86 & F1amplitudeLogRelF0\_sma3nz\_amean & 0.76 \\
87 & StddevVoicedSegmentLengthSec & 0.72 \\
88 & F0semitoneFrom27.5Hz\_sma3nz\_stddevRisingSlope & 0.67 \\
89 & StddevUnvoicedSegmentLength & 0.67 \\
90 & F0semitoneFrom27.5Hz\_sma3nz\_stddevFallingSlope & 0.65 \\
\noalign{\smallskip}\hline
\end{tabular}
\end{table}

\end{document}